\documentclass{elsart}

\usepackage{graphicx}
\usepackage{amssymb}
\usepackage{epsfig}

\def\Journal#1#2#3#4{{#1} {#2} (#3) #4}

\def\NPA{Nucl. Phys. A}
\def\PLB{Phys. Lett.  B}

\def\be{\begin{equation}}
\def\ee{\end{equation}}

\newcommand{\ud}{\mathrm{d}}

\begin{document}
\begin{frontmatter}

\title{Charmonium and open charm production in nuclear collisions at SPS/FAIR
  energies and the possible influence of a hot hadronic medium}

\author[gsi]{A.~Andronic}, 
\author[gsi,tud]{P.~Braun-Munzinger},
\author[wro,tud]{K.~Redlich},
\author[hei]{J.~Stachel}

\address[gsi]{Gesellschaft f\"ur Schwerionenforschung, GSI, 
D-64291 Darmstadt, Germany}
\address[tud]{Technical University Darmstadt, D-64289 Darmstadt, Germany}
\address[wro]{Institute of Theoretical Physics, University of Wroc\l aw,
PL-50204 Wroc\l aw, Poland}
\address[hei]{Physikalisches Institut der Universit\"at Heidelberg,
D-69120 Heidelberg, Germany}

\begin{abstract} 
  We provide predictions for charmonium and open charm production in nuclear
  collisions at SPS/FAIR energies within the framework of the statistical
  hadronization model. The increasing importance at lower energies of
  $\Lambda_c$ production is demonstrated and provides a challenge for future
  experiments. We also demonstrate that, because of the large charm quark mass
  and the different timescales for charm quark and charmed hadron production,
  possible modifications of charmed hadrons in the hot hadronic medium do not
  lead to measurable changes in cross sections for D-meson production. A
  possible influence of medium effects can be seen, however, in yields of
  charmonium. These effects are visible at all energies
  and  results are presented for the energy range between charm threshold
  and RHIC energy. 
\end{abstract}

\vspace{2mm}

\end{frontmatter}

\section{Introduction}

Charmonium production is considered, since the original proposal more than 20
years ago about its suppression in a Quark-Gluon Plasma (QGP)
\cite{satz}, as an important probe to determine the degree of deconfinement
reached in the fireball produced in ultra-relativistic nucleus-nucleus
collisions. In a recent series of publications \cite{aa1,aa2,aa3} we have
demonstrated that, in the energy range from top SPS energy ($\sqrt{s_{NN}}
\approx 17$ GeV) on,  the data on $J/\psi$ and $\psi'$ production in
nucleus-nucleus collisions can be well described within the statistical
hadronization model proposed in \cite{pbm1}. In particular the, at first
glance surprizing, rapidity dependence of the nuclear modification factor
$R_{AA}^{J/\psi}$ observed by the PHENIX collaboration \cite{phe1} is 
explained as due to the statistical hadronization
of c {and $\bar c$ pairs at 
the phase boundary between quark gluon plasma and hot hadronic 
matter.\footnote{$R_{AA}^{J/\psi}$ is defined as 
$R_{AA}^{J/\psi}= \frac{\ud N_{J/\psi}^{AuAu}/\ud
    y}{N_{coll}\cdot\ud N_{J/\psi}^{pp}/\ud y}$ 
and relates the charmonium yield in nucleus-nucleus collisions to that
expected for a superposition of independent nucleon-nucleon collisions.
In this expression, $\ud N_{J/\psi}/\ud y$ is the rapidity density of the
$J/\psi$ yield integrated over transverse momentum and $N_{coll}$ is the
number of binary collisions for a given centrality class.}
We note that extrapolation to LHC energy of these results yields \cite{aa2,aa3}
a rather striking centrality dependence and, depending on the magnitude of 
the $\bar c c$ cross section in central Pb-Pb collisions, possibly even 
an enhancement ($R_{AA}^{J/\psi} > 1$)  of $J/\psi$ production due to 
hadronization of thermalized, deconfined and in general uncorrelated charm 
quarks.

In the present publication we explore, for the first time, the lower energy
range from near threshold ($\sqrt{s_{NN}} \approx 6$ GeV) to top SPS energy.
The lower part of this  energy range can be investigated in the CBM experiment
\cite{cbm1}  
at the future FAIR facility. 
One of the motivations for such studies was the expectation\cite{cbm1,tol} 
to provide, by a measurement of D-meson production near threshold, 
information on their possible modification near the phase boundary 
. Here we demonstrate that, because of the relevant mass 
and time scales involved, medium effects on D-meson production are likely 
to be very small. Furthermore, because of the dominance of 
baryochemical potential (coupled with the charm neutrality condition) 
at low energies, it turns out to be important to measure
in addition to D-meson production also the yield of charmed baryons  
to get a good measure of the total charm production cross section.

In section 2 we will discuss the various time scales relevant for charm,
charmonium, and open charm hadron production and discuss their relevance for
the applicability of the statistical hadronization model as well as for
the study of possible medium effects in the charm sector. Section 3 will
provide a brief review of the statistical hadronization model. Our results 
on open charm and charmonium production from  low beam energies on will be 
presented in section 4. In section 5 we will introduce various possible
medium effects on open charm hadrons and study their influence on measurable
quantities from FAIR to RHIC energies before concluding with a brief 
outlook in the last section.

\section{On relevant time scales and medium effects}

In the original scenario of $J/\psi$ suppression via Debye screening
\cite{satz} it is assumed that the charmonia are rapidly formed in initial
hard collisions but are subsequently destroyed in the QGP. While it is clear
that the production of a (colored)  charm quark pair takes place at
time $t_{c \bar c} = 1/(2m_c) \leq 0.1$ fm, the formation time of charmonium
involves color neutralization and the build-up of its wave function. The
relevant time scale has been studied early on \cite{kar,bla}
and is of order 1 fm. Similar arguments also apply for the production time of
charmed hadrons and we expect comparable time scales as for charmonium. 

We note that, at SPS energy where the 'melting scenario' was originally 
studied, this time is in the same range as the plasma formation
time. At SPS and lower energies, charmonia can be formed in the pre-plasma
phase and must be destroyed in the plasma if suppression by QGP is to take
place. 

At the collider energies of RHIC and especially LHC the plasma formation time
is likely to be much shorter (comparable to $t_{c \bar c}$).  Furthermore, the
number of charm quark pairs can exceed 10 per unit rapidity (central collisions
at LHC). Initially, the 'collider' plasma will be hotter than T$_D$, the 
temperature above which screening takes place, and no charmonia will be 
formed at all in the QGP. 
It is our view that the charm quarks will be effectively thermalized leading 
to an uncorrelated pool of c and $\bar c$ quarks.  Once the plasma temperature
falls below T$_D$ charmonia can be formed in principle, 
as well as destroyed, but as is indicated by the studies performed in 
\cite{aa2}, their formation rate is likely to be low.
This finally leads to the notion, expressed explicitely in the statistical
hadronization model, that all charmonia are produced by (re-)combination of
charm quarks at the phase boundary. We would like to emphasize that, in this
scenario, the particular value of T$_D$ which is much discussed in the recent
literature \cite{satz2,moc}, is not very important.  Models that combine the
'melting scenario' with statistical hadronization have been proposed
\cite{gra0}. Alternatively, charmonium formation by coalescence in the plasma
\cite{the1,the2,gra,yan} as well as within transport model approaches
\cite{zha,bra} has been considered.

Another issue to be considered is the collision time
$t_{coll}=2R/\gamma_{cm}$, where $R$ is the radius of the (assumed equal mass)
nuclei and $\gamma_{cm}$ is the Lorentz $\gamma$ factor of each of the beams
in the center-of-mass system.  At SPS and lower energies, $\gamma_{cm} < 10$
and t$_{coll} > 1$ fm for a central Au-Au or Pb-Pb collision, so collision
time, plasma formation time, and charmonium formation time are all of the same
order \cite{bla}. Furthermore, the maximum plasma temperature may not exceed 
T$_D$. In this situation the formed charmonia may be broken up 
by gluons and by the high energy nucleons still passing by from the 
collision. In this latter case, cold nuclear suppression needs to be 
carefully considered, as discussed, e.g., in \cite{satz1,arleo}.
However, we note in this context that the charm quarks resulting from such
break-up processes eventually have to hadronize, which might again lead to
charmonium production at the phase boundary if the charm quarks are
kinetically thermalized, as is assumed in the statistical hadronization model
\cite{pbm1,aa2}. Consequently, our calculations, in which both charmonium
formation before QGP production and cold nuclear suppression are neglected,
may  somewhat underestimate the charmonium production yield at SPS energies
and below. In that sense the below calculated medium effects are upper limits
for energies close to threshold.

At collider energies there will be yet another separation of time scales. 
At LHC energy, the momentum of a Pb nucleus
is $p_{cm}$=2.76 TeV per nucleon, leading to $\gamma_{cm}=2940$, hence 
$t_{coll} < 5 \cdot 10^{-3}$ fm. Even ``wee'' partons with
momentum fraction\footnote{We choose this value since it corresponds to a wee
  parton energy of approximately the binding energy of the $J/\psi$ meson.
  Smaller x values are hence not relevant for the present considerations.}
$x_w = 2.5 \cdot 10^{-4}$ will pass by within a time $t_w = 1/(x p_{cm}) <
0.3$ fm, and will not destroy any charmonia since none exist at that time. We
consequently expect that cold nuclear absorption will decrease from SPS to
RHIC energy and should be negligible at LHC energy. First indications for this
trend are visible in the PHENIX data \cite{phe0}.

Given the various time scales it becomes clear from the above discussion that
the statistical hadronization model should become a quantitative tool to
describe charmonium and open charm production at collider energies without the
explicit need to take account of any charmonium or open charm hadron formation
before the QGP is developed and of cold nuclear absorption effects.  We note
in passing that the issue of shadowing or saturation effects is of an entirely
different nature: within the framework of the statistical hadronization model
we need to know the rapidity density for open charm production in
nucleus-nucleus collisions. Using this quantity, which of course contains
shadowing or saturation effects, as input we can then provide cross sections
for the production of all open and hidden charm hadrons.

Finally we would like to discuss the effect of possible in-medium changes of
charmed hadrons on their production cross section. We start the discussion by
recalling that 
\be \sigma_{c \bar c} = \frac{1}{2} ( \sigma_D +
\sigma_{\Lambda_c} +\sigma_{\Xi_c} + ...) + ( \sigma_{\eta_c} +
\sigma_{J/\psi} + \sigma_{\chi_c} + ...)
\label{aa_eq0}  
\ee because of charm conservation.
As shown in \cite{aa2}, annihilation of charm quarks can be fully neglected.
In the above equation, $\sigma_D$ is the total cross section for the 
production of any D-meson. The cross section $\sigma_{c \bar c}$ is governed 
by the mass of the charm quark $m_c \approx 1.3$ GeV \cite{pdg}, which is 
much larger  than any soft Quantum Chromodynamics (QCD) scale such as 
$\Lambda_{QCD}$. 
Therefore we expect no medium effects on this quantity. 
Such a separation of scales is not possible for strangeness production, and
the situation there is not easily comparable. 

The much later formed D-mesons, or other charmed hadrons, may well change
their mass in the hot medium. The results of various studies on in-medium
modification of charmed hadrons masses
\cite{tol,tsu,sib1,sib,hay,cas,fri,lutz}  are sometimes contradictory.
Within a QCD sum rule model, Hayashigaki \cite{hay} predicts for $D$ mesons a
50 MeV mass decrease at normal nuclear density, while for $J/\psi$ meson the
shift is much smaller (5 MeV).  Friman et al. \cite{fri} have concluded that
the widths are little affected in dropping mass scenarios, while Tolos et al.
\cite{tol} assert that only the widths could be affected by the nuclear 
medium but not the (pole) mass of $D$ mesons.  It is not clear whether 
the mass changes are different for particles and antiparticles, as advocated
in \cite{sib1}, or identical because the vector potential may not matter
if one assumes production at an early stage \cite{cas}. 
The "indirect" effect on $J/\psi$ production was also investigated 
\cite{sib,hay,zha,fri,gra}.  A large mass shift of $J/\psi$ has been recently 
advocated \cite{mor}.
We note that excellent fits of the common (non-charmed) hadrons to 
predictions of the thermal model have been obtained using vacuum masses 
\cite{pbm0,pbm0a,aat}. An attempt to use modified masses for the
RHIC energy \cite{bro} has not produced a conclusive preference
for any mass or width modifications of hadrons in medium. On the other hand,
some evidence for possible mass modifications was presented in the chiral
model of \cite{zschiesche}.
   
Whatever the medium effects may be, they can, because of the charm 
conservation equation above, only lead to a redistribution of charm quarks.
In particular, if the masses of D-mesons  are lowered by the same amount 
at the phase boundary, this effect would practically not be visible in 
the D-meson cross section.  Although the charm conservation equation above 
is strictly correct only for the total cross section we expect within the 
framework of the statistical hadronization model, also little influence 
due to medium effects on distributions in rapidity and transverse momentum.
This is due to the fact that the crucial input into our model is 
$\ud N_{c\bar{c}}^{AuAu}/\ud y$ and there is no substantial D-meson 
rescattering after formation at the phase boundary.
Modification of D-meson masses at the phase boundary will, however, 
influence the production rates for charmonia: after lowering of their masses
the D-mesons will eat away the charm quarks of the charmonia but since the
D-mesons are much more abundant, their own yield will hardly change 
because of total charm conservation. 

Much of the above argument about medium effects is essentially
model-independent and applies equally well at all energies.  Here we will
consider various types of scenarios for medium modifications and study their
effect within the statistical hadronization framework in the energy range 
from charm threshold to collider energies.

\section{Reminder of ingredients and assumptions of the statistical
  hadronization model} 

The statistical hadronization model (SHM) \cite{pbm1,aa2} assumes that 
the charm quarks are produced in primary hard collisions and
that their total number stays constant until hadronization.  
Another important factor is thermal equilibration in the QGP, at least 
near the critical temperature, $T_c$. 
While data at RHIC energy \cite{phe2} suggest charm equilibration,
at lower energies, where the initial densities and temperatures
are lower, the assumption of equilibration can be questionable.
In this exploratory study we have nevertheless assumed full thermalization.

We focus on the energy range $\sqrt{s_{NN}}$=7-200 GeV and perform 
calculations for central collisions of heavy equal mass nuclei, 
corresponding to $N_{part}$=350.
We neglect charmonium production in the nuclear corona  \cite{aa2},  since we
focus in the following on central collisions, where such effects are small. 

In the following we briefly outline the calculation steps in our model
\cite{pbm1,aa2}.  
The model has the following input parameters:
i) charm production cross section in pp collisions;
ii) characteristics at chemical freeze-out: temperature, $T$, 
baryochemical potential, $\mu_b$, and volume corresponding to one unit 
of rapidity $V_{\Delta y=1}$ (our calculations are for midrapidity). Since, in
the end, our main results will be ratios of hadrons with charm quarks
nomalized to the $\bar c c$ yield, the detailed magnitude of the open charm
cross section and whether to use integrated yield or midrapidity yields
is not crucial.

The charm balance equation \cite{pbm1}, which has to include canonical 
suppression factors \cite{gor} whenever the number of charm pairs is 
not much larger than 1, is used to determine a fugacity  factor $g_c$ via:
\begin{equation}
N_{c\bar{c}}^{dir}=\frac{1}{2}g_c N_{oc}^{th}
\frac{I_1(g_cN_{oc}^{th})}{I_0(g_cN_{oc}^{th})} + g_c^2N_{c\bar c}^{th}.
\label{aa:eq1}
\end{equation}
Here $N_{c\bar{c}}^{dir}$  is the number of initially produced $c\bar{c}$ 
pairs and  $I_n$ are  modified Bessel functions. In the fireball of volume 
$V$ the total number of open ($N_{oc}^{th}=n_{oc}^{th}V$) and hidden 
($N_{c\bar c}^{th}=n_{c\bar c}^{th}V$) charm hadrons is computed from 
their grand-canonical densities $n_{oc}^{th}$ and $n_{c\bar c}^{th}$, 
respectively. This charm balance equation is the implementation within our
model of the charm conservation constraint expressed in eq.\ref{aa_eq0}.
The densities of different particle species in the grand canonical ensemble 
are calculated following the statistical model \cite{pbm0,pbm0a,aat}.
The balance equation (\ref{aa:eq1}) defines the fugacity parameter $g_c$ 
that accounts for deviations of heavy quark multiplicity from the value 
that is expected in complete chemical equilibrium. 
The yield of charmonia of type $j$ is obtained as: $N_j=g_c^2 N_j^{th}$,
while the yield of open charm hadrons is: 
$N_i=g_c N_i^{th}{I_1(g_cN_{oc}^{th})}/{I_0(g_cN_{oc}^{th})}$.

\begin{figure}[ht]
\vspace{-.7cm}
\begin{tabular}{lc} \begin{minipage}{.62\textwidth}
\hspace{-.4cm}\includegraphics[width=1.2\textwidth,height=1.13\textwidth]{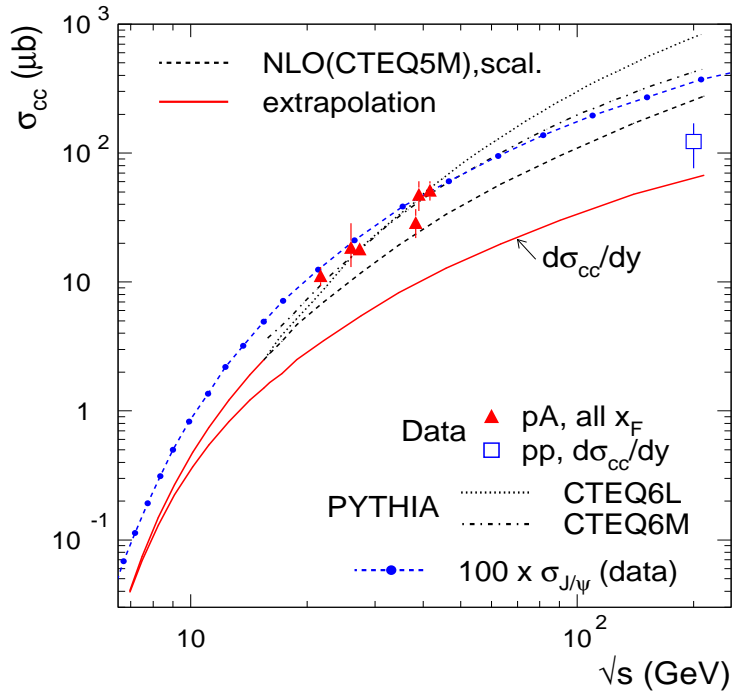}
\end{minipage}  & \begin{minipage}{.34\textwidth}
\caption{Energy dependence of the charm production cross 
section in pp collisions. The NLO pQCD values \cite{rv1} are 
compared to calculations using PYTHIA and to data in pA collisions, taken 
from ref. \cite{lou}. 
Our extrapolations for low energies are shown with continuous lines, 
for total and midrapidity ($\ud\sigma_{c\bar{c}}/\ud y$) cross section.
The open square is a midrapidity measurement in pp collisions \cite{phe3}.
The dashed line with dots indicates a parameterization of the measured energy
dependence of the $J/\psi$ production cross section \cite{herab}.} 
\label{aa_fig0}
\end{minipage} \end{tabular} 
\end{figure}

As no information on the charm production cross section is available for 
energies below $\sqrt{s}$=15 GeV, we have to rely on extrapolation. 
The basis for this extrapolation is the energy dependence of the total 
charm production cross section calculated in ref. \cite{rv1} for the 
CTEQ5M parton distribution functions in next-to-leading order (NLO), 
as shown in Fig.~\ref{aa_fig0}.
We have scaled these calculations to match the more recent values
calculated at $\sqrt{s}$=200 GeV in ref. \cite{cac}.
We employ a threshold-based extrapolation using the following expression:
\begin{equation}
\sigma_{c\bar{c}}=k (1-\sqrt{s_{thr}}/\sqrt{s})^a(\sqrt{s_{thr}}/\sqrt{s})^b
\end{equation}
with $k$=1.85 $\mu$b, $\sqrt{s_{thr}}$=4.5 GeV (calculated assuming a charm
quark mass $m_c$=1.3 GeV), $a$=4.3, and $b$=-1.44. The parameters $a$, $b$,
$k$ were tuned to reproduce the low-energy part of the (scaled) NLO curve.
The extrapolated curves for charm production cross section are shown with
continuous lines in Fig.~\ref{aa_fig0}.  Also shown for comparison are
calculations with PYTHIA \cite{lou}.  To obtain the values at midrapidity we
have extrapolated to lower energies the rapidity widths (FWHM) of the charm
cross section known to be about 4 units at RHIC \cite{cac} and about 2 units
at SPS \cite{pbm2}.  With these cross section values, the rapidity density of
initially produced charm quark pairs strongly rises from 1.1$\cdot$10$^{-3}$
to 1.7 for the energy range $\sqrt{s_{NN}}$=7-200 GeV.  We note that the
so-obtained charm production cross section has an energy dependence similar to
that measured for $J/\psi$ production, recently compiled and parametrized by
the HERA-B collaboration \cite{herab}.  For comparison, this is also shown in
Fig.~\ref{aa_fig0}.  The extrapolation procedure for the low-energy part of
the cross section obviously implies significant uncertainties. We emphasize,
however, that the most robust predictions of our model, i.e. the yields of
charmed hadrons and charmonia relative to the initially produced $c \bar c$
pair yield are not influenced by the details of this extrapolation.

For the studied energy range, $T$ rises from 151 to 161 MeV from
$\sqrt{s_{NN}}$=7 to 12 GeV and stays constant for higher energies, while
$\mu_b$ decreases from 434 to 22 MeV \cite{aat}.  The volume $V_{\Delta y=1}$
at midrapidity continuously rises \cite{aat} from 760 to 2400 fm$^3$.  Due to
the very small number of charm quarks at these low energies, the canonical
suppression factor ($I_1/I_0$) is very large, but strongly decreases from
about 1/30 to 1/1.2 for $\sqrt{s_{NN}}$=7 to 200 GeV. Correspondingly, the
charm fugacity $g_c$ increases from 0.96 to 8.9.

\section{Energy dependence of charmed hadrons yield}

Our main results are presented in Fig.~\ref{aa_fig1}.  The upper panel shows
our predictions for the energy dependence of midrapidity yields for various
charmed hadrons.  Beyond the generally decreasing trend towards low energies
for all yields one notices first a striking behavior of the production of 
$\Lambda_c$ baryons: their yield exhibits a weaker energy dependence than 
observed for other charmed hadrons. In our approach this is caused by the 
increase in baryochemical potential towards lower energies.  A similar 
behavior is seen for the $\Xi_c^+$ baryon.  In detail, the production yields
of D-mesons depend also on their quark content.

\begin{figure}[hbt]
\centering\includegraphics[width=.6\textwidth]{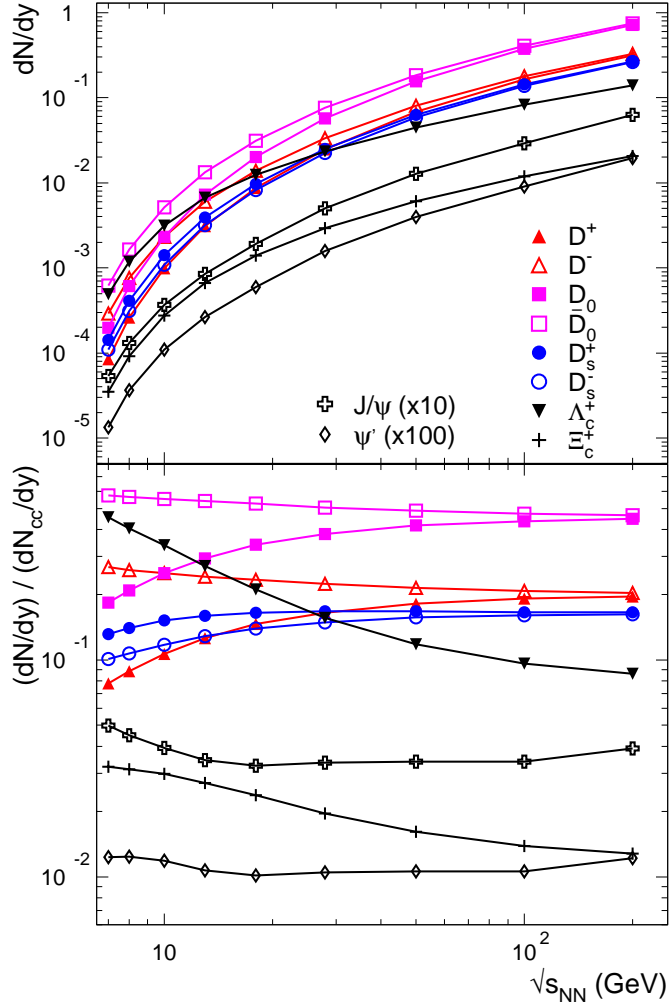}
\caption{Energy dependence of charmed hadron production at midrapidity.
Upper panel: absolute yields, lower panel: yields relative to the number
of $c\bar{c}$ pairs. Note the scale factors of 10 and 100 for $J/\psi$ 
and $\psi'$ mesons, respectively.} 
\label{aa_fig1}
\end{figure} 

The differing energy dependences of the yields of charmed hadrons
are even more evident in the lower panel of Fig.~\ref{aa_fig1}, where 
we show the predicted yields normalized to the number of initially produced
$c\bar{c}$ pairs.
Except very near threshold, the $J/\psi$ production 
yield per $c\bar{c}$ pair exhibits a slow increase with increasing energy. 
This increase is a consequence of the quadratic term in the $J/\psi$ yield 
equation discussed above. 
At LHC energy, the yield ratio $J/\psi/c\bar{c}$ approaches 
1\% \cite{aa2}, scaling linearly with $\sigma_{c\bar{c}}$; for details see
\cite{aa4}. 
The $\psi'$ yield shows a similar energy dependence as the $J/\psi$,
except for our lowest energies, where the difference is due to the
decrease of temperature (see above).
We emphasize again that this model prediction, namely yields relative to 
$c\bar{c}$ pairs, is a robust result, as it is in the first order independent 
on the charm production cross section.
Due to the expected similar temperature, the relative abundance of open charm 
hadrons at LHC is predicted \cite{aa4} to be similar to that at RHIC energies.

\section{Effects of in-medium modification of charmed hadrons masses}

\begin{figure}[htb]
\centering\includegraphics[width=1.0\textwidth]{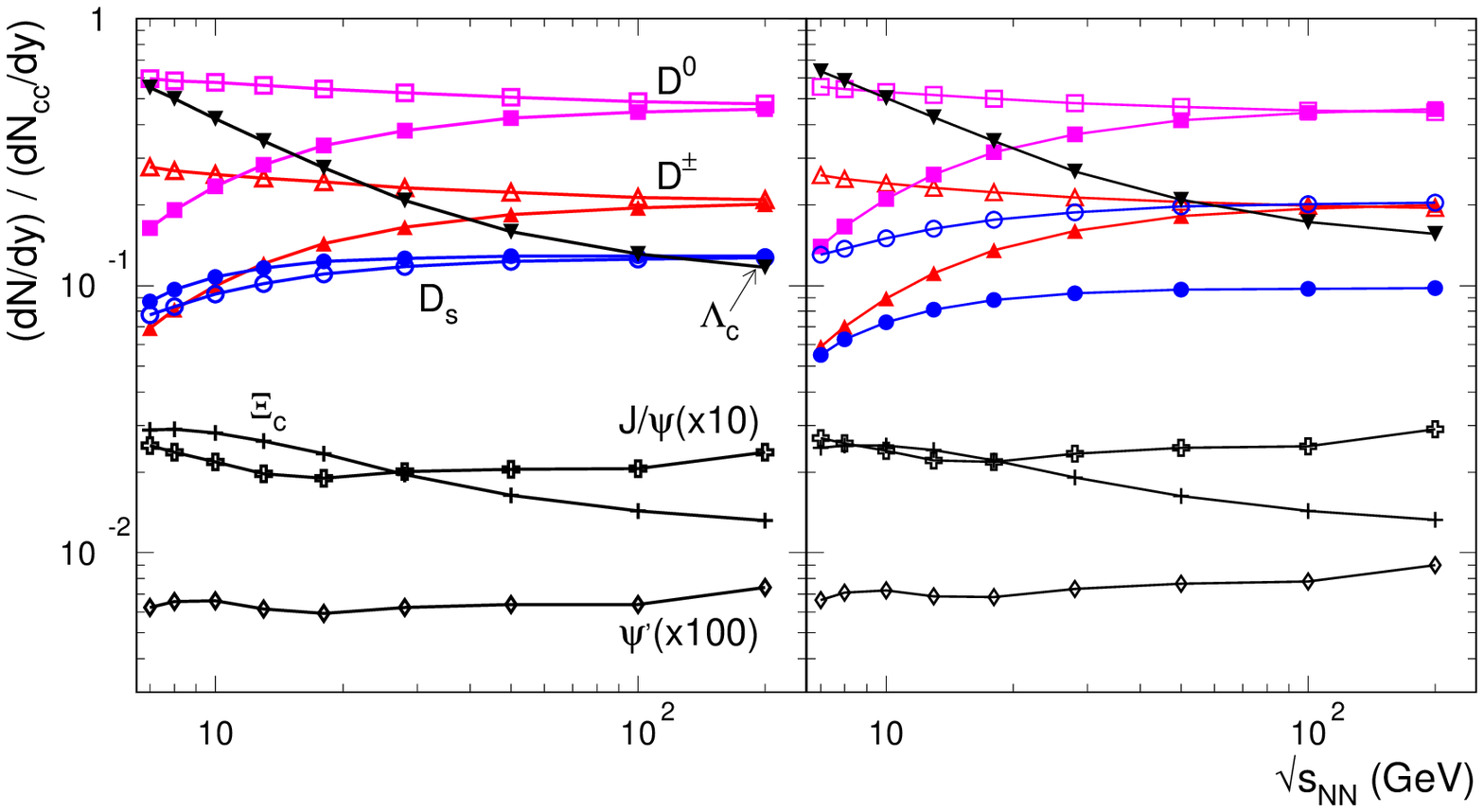}
\caption{Energy dependence of the yield of charmed hadrons relative to
the charm quark pair yield for two scenarios of the mass change 
(left panel for scenario i), right panel for scenario ii), see text).
For the D mesons, the full and open symbols are for particles and 
antiparticles, respectively. Note the factors 10 and 100
for the $J/\psi$ and $\psi'$ mesons, respectively.} 
\label{aa_fig2x}
\end{figure} 

We consider two scenarios\footnote{The scenarios are constructed by
  modification of the constituent quark masses of light ($u$ and $d$) quarks in
  the charmed hadrons by fixed amounts. Reducing, for example, the light quark
  masses by 50 MeV will lower D-meson masses by 50 MeV and the $\Lambda_c
  (\Xi_c)$ mass by 100 (50) MeV.} for a possible mass change $\Delta m$ of open
charm hadrons containing light, $u$ or $d$, quarks: i) a common decrease of 50
MeV for all charmed mesons and their antiparticles and a decrease of 100 MeV
for the $\Lambda_c$ and $\Sigma_c$ baryons (50 MeV decrease for $\Xi_c$); ii)
a decrease of 100 MeV for all charmed mesons and a 50 MeV increase for their
antiparticles, with the same (scaled with the number of light quarks) scenario
as in i) for the baryons.  Scenario i) is more suited for an isospin-symmetric
fireball produced in high-energy collisions and was used in \cite{cas}, while
scenario ii) may be realized at low energies.  In both scenarios, the masses
of the $D_s$ mesons and of the charmonia are the vacuum masses. 
We also note that if one leaves all D-meson masses unchanged but 
allows their widths to increase, the resulting yields will increase by 11\%
(2.7\%) for a  width of 100 MeV (50 MeV).
If the in-medium widths exhibit tails towards low masses, as has been 
suggested by \cite{tol}, to first order the effect on thermal densities
is comparable with that from a decrease in the pole mass.

The results for the two cases are presented in Fig.~\ref{aa_fig2x} as yields
relative to the number of initially-produced $c\bar{c}$ pairs.  As a result of
the redistribution of the charm quarks over the various species, the relative
yields of charmed hadrons may change. For example, in scenario i) the ratios 
of D-mesons are all close to those computed for vacuum masses
(Fig.~\ref{aa_fig1}), but the $\Lambda_c$/D ratio is increased.  Obvious are
for scenario ii) the changes in the relative abundances of the $D$ and
$\bar{D}$ mesons, as well as for the charmed baryons and also the relative
production yields of $D^+_s$ and $D^-_s$ mesons are very different. This
difference is the result of the asymmetry in the mass shifts for particles 
and antiparticles.
The change in yield of $D^\pm _s$ mesons occur as a consequence of the charm
neutrality condition. Overall, however, charm conservation leads to rather 
small changes in the total yields. In contra-distinction, the effect of mass 
changes of charmed hadrons is very significant for charmonia, in particular 
the yields are more affected at low energies.

\begin{figure}[hbt]
\vspace{-1.cm}
\centering\includegraphics[width=.73\textwidth]{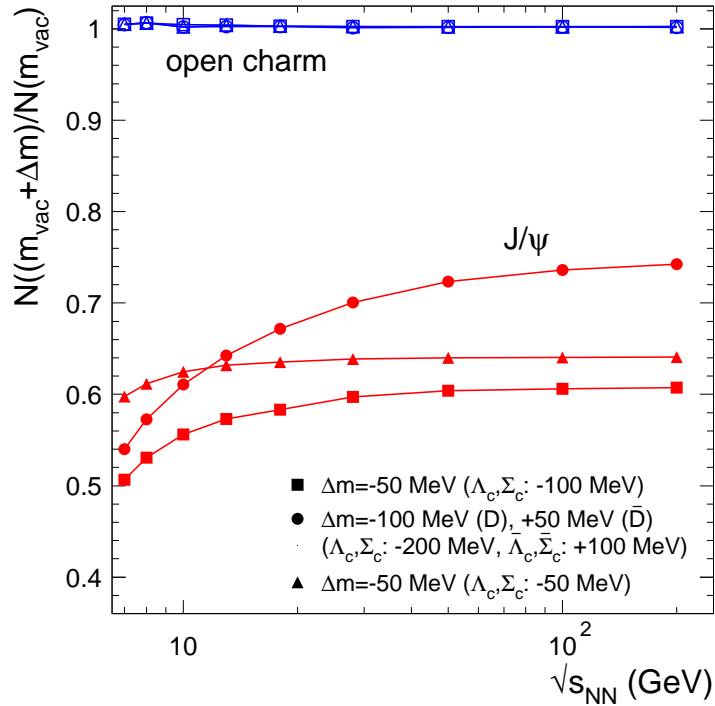}
\caption{Energy dependence of the relative change in the production
yield of open charm hadrons and of $J/\psi$ meson considering 
different scenarios for in-medium mass modifications (see text).} 
\label{aa_fig2}
\end{figure} 

In Fig.~\ref{aa_fig2} we demonstrate that the total open charm yield (sum over
all charmed hadrons) exhibits essentially no change if one considers mass
shifts, while the effect is large on charmonia. This is to be expected from
eq. \ref{aa:eq1}: as D-meson and $\Lambda_c$-baryon masses are reduced, e.g.,
the charm fugacity $g_c$ is changed accordingly to conserve charm.  Since the
D-meson and $\Lambda_c$-baryon yield varies linearly with $g_c$ we expect
little change, while yields for charmonia vary strongly, since their yields
are proportional to $g_c^2$.  To demonstrate this we plot, in
Fig.~\ref{aa_fig2}, the relative change of the yields with in-medium masses
compared to the vacuum case. For this comparison, we have added a third case,
namely considering that the mass change of charmed baryons is the same as for
the mesons.

Despite showing similar trends, in our model the reduction of the $J/\psi$ 
yield due to reduced in-medium masses of open charm hadrons has a different 
origin than that investigated in previous studies \cite{sib,hay,zha,fri,gra}.
Dissociation of $J/\psi$ in a gas of $\pi$ and $\rho$ mesons \cite{sib} and
effects originating from opening of decay channels of $\psi'$ and $\chi_c$ 
states into $D\bar{D}$ \cite{fri,gra} were considered in hadronic scenarios,
while Zhang et al. \cite{zha} have investigated the effect of mass changes
both in the partonic ($c$ quarks) and in the hadronic (D mesons) stage.
In our model all hadrons with open and hidden charm are produced at chemical
freeze-out. Under the assumption that chemical freeze-out takes place at 
the phase boundary \cite{pbm3}, the medium effects could be due to the onset 
of chiral symmetry restoration or rescattering with the constituents of 
the medium.
At present it cannot be ruled out that at lower energies the phase boundary 
does not coincide with chemical freeze-out. In this case, in-medium effects 
would be due to rescattering in the dense hadronic phase.

\section{Conclusions}

We have investigated charmonium production in the statistical hadronization
model at lower energies. An interesting result is that the yield of charmed
baryons ($\Lambda_c$, $\Xi_c$) relative to the total $c\bar{c}$ yield
increases strongly with decreasing energy. Below $\sqrt{s_{NN}}$=10 GeV, the
relative yield of $\Lambda_c$ exceeds that of any D meson except $\bar{D}_0$,
implying that an investigation of open charm production at low energies needs
to include careful measurements of charmed baryons, a difficult experimental task.
The charmonium/open charm yield rises only slowly from energies near threshold 
to reach $\sim$1\% at LHC energy.  Note that this ratio depends on the magnitude
of the charm cross section, further underlining the importance to measure this
quantity with precision.  
Our study is the first one addressing comprehensively the charm redistribution
in principle and under various assumptions of in-medium masses of charmed 
hadrons. 
Because of a separation of time scales for charm quark and charmed hadron
production, the overall D meson cross section is very little affected by
in-medium mass changes, if charm conservation is taken into account.
Measurable effects are predicted for the yields of charmonia. These effects
are visible at all beam energies and increase slightly towards threshold.

\section*{Acknowledgments}
K.R. acknowledges partial support from the Polish Ministry of Science (MEN)
and the Deutsche Forschungsgemeinschaft (DFG) under Mercator Programme.

\end{document}